\documentclass[aps,prl,reprint,superscriptaddress,letterpaper]{revtex4-1}
\usepackage[english]{babel}
\usepackage{ucs}
\usepackage[utf8x]{inputenc}
\usepackage{graphicx}
\usepackage{xcolor}
\usepackage{transparent}
\usepackage{bm}
\usepackage{bbm}
\usepackage{empheq}
\newcommand{\bra}[1]{\langle #1|}
\newcommand{\ket}[1]{|#1\rangle}

\newcommand{\mm}[1]{\mathrm{#1}}
\newcommand{\ui}{\mathrm{i}}
\newcommand{\ue}{\mathrm{e}}
\newcommand{\uh}{\mathrm{h}}

\newcommand{\Bv}{\mbox{\boldmath$B$}}

\newcommand{\abs}[1]{\left|#1\right|}

\begin{document}

\title{Nuclear Spin Diffusion Mediated by Heavy Hole Hyperfine Non-Collinear Interactions}

\author{Hugo Ribeiro}
\affiliation{Department of Physics, University of Basel, Klingelbergstrasse 82, CH-4056
Basel, Switzerland}

\author{Franziska Maier}
\affiliation{Department of Physics, University of Basel, Klingelbergstrasse 82, CH-4056
Basel, Switzerland}

\author{Daniel Loss}
\affiliation{Department of Physics, University of Basel, Klingelbergstrasse 82, CH-4056
Basel, Switzerland}

\date{\today}

\begin{abstract}
We show that the hyperfine mediated dynamics of heavy hole states confined in neutral
self-assembled quantum dots leads to a nuclear spin diffusion mechanism. It is found
that the oftentimes neglected effective heavy hole hyperfine non-collinear interaction is
responsible for the low degree of nuclear spin polarization in neutral quantum dots.
Moreover, our results demonstrate that after pumping the nuclear spin state is left in a
complex mixed state, from which it is not straightforward to deduce the sign of the
Ising-like interactions.
\end{abstract}

\maketitle

We are currently in the midst of an effort to develop reliable nanostructures that can be
used to host qubits. Among the possible architectures~\cite{monroe2013, devoret2013,
awschalom2013, stern2013}, the progress made with spin-based qubits confined in
semiconductor structures~\cite{loss1998} has been the most
impressive~\cite{kloeffel2013}. In only a decade, it became possible to efficiently
initialize~\cite{ono2002}, manipulate coherently~\cite{petta2005, koppens2006,
nowack2007, berezovsky2008, press2008, petta2010, ribeiro2013}, and measure the state of
a single spin confined in both electrically defined and self-assembled quantum dots.  All
of these remarkable achievements are, however, mitigated by poor coherence times on the
order of tens of nanoseconds~\cite{petta2005, koppens2006, nowack2007, berezovsky2008}.
In quantum dots made out of III-V materials, the fluctuations of the nuclear spin felt by
the electronic spin through the hyperfine interaction are the main source of
decoherence~\cite{petta2005, koppens2006, khaetskii2002, merkulov2002, coish2004,
coish2005, coish2009, xu2009}. Nevertheless, dynamical decoupling schemes have improved
the situation and revealed longer dephasing times~\cite{greilich2009, clark2009,
press2010, bluhm2011}.

From another perspective, nuclear spins are a helpful resource for quantum computing.
In gate defined dots, coherent manipulation of electron spin states via the hyperfine
interaction has been demonstrated~\cite{petta2010,ribeiro2010,gaudreau2012}. In
self-assembled dots, direct control of nuclear spins has been realized via nuclear
magnetic resonance (NMR)~\cite{makhonin2011,munsch2013}, which allows to control the
direction of the Overhauser field and consequently can be used to control an electron
(hole) spin-based qubit. 
In spite of the efforts made to harness nuclear spins, the role of heavy holes in
the dynamics is not yet fully understood.

The first theories suggested an Ising-like type of interaction with a strength on the
order of $10\%$ of the one of the electron and with opposite sign~\cite{fischer2008,
testelin2009}, which was experimentally verified~\cite{fallahi2010, chekhovich2011}.
However, subsequent experiments seem to contradict these early results. It has recently
been claimed that the sign of the coupling strength is opposite for cations and
anions~\cite{chekhovich2013}. Some other recent experiments~\cite{xu2009, carter2013}
report results which indicate a feedback mechanism between heavy holes and nuclear
spins. Theories based on $p$-symmetric Bloch functions for hole states predict that
flip-flop terms similar to those of the electronic hyperfine Hamiltonian are very
weak~\cite{fischer2008, testelin2009}. Consequently, it was proposed that non-collinear
hyperfine interactions could account for the joint heavy hole nuclear spin
dynamics~\cite{yang2012, shi2013}. However, non-collinear interactions were predicted to
only affect the dynamics if the laser frequency is not on resonance with the electronic
transition which is being driven. An alternative explanation would be that hole states
have to be described by both $p$- and $d$-type Bloch functions~\cite{chekhovich2013}
leading to a stronger flip-flop exchange mechanism.

\begin{figure}
\includegraphics[width=\columnwidth]{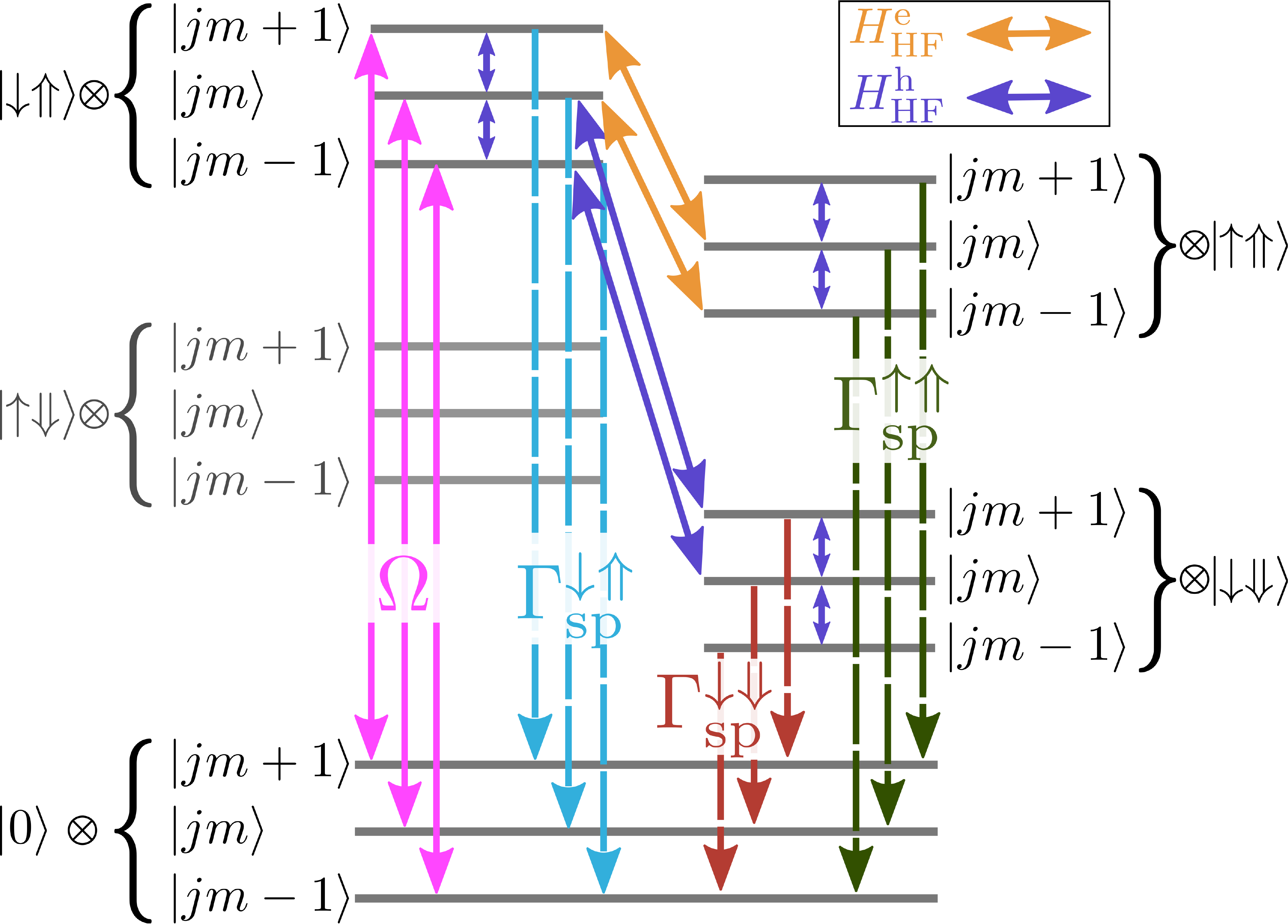}
\caption{(Color online). Level scheme of the excitonic states in a neutral quantum dot
showing optically driven transitions with Rabi frequency $\Omega$ under the absorption of
$\sigma_+$ polarized light (magenta). The nuclear states are described with the total
angular momentum $j$ and magnetization $m$. Hyperfine mediated transitions via the
electron are shown in orange and in purple for the hole spin. The excited states relax
via spontaneous emission with rates $\Gamma_{\mm{sp}}^{\downarrow \Uparrow} \approx
\Gamma_{\mm{sp}}^{\uparrow \Downarrow} \gg \Gamma_{\mm{sp}}^{\uparrow \Uparrow} \approx
\Gamma_{\mm{sp}}^{\downarrow \Downarrow}$.}
\label{fig:opticalprocesses}
\end{figure}

In this letter, by focusing on optical pumping of nuclear spins in neutral quantum dots,
we show that the effective hyperfine interaction for heavy hole states, described with
$p$-symmetric Bloch functions, via the non-collinear term leads to an effective nuclear
spin diffusion mechanism. Opposite to earlier theories, we find that non-collinear
interactions influence the nuclear spin dynamics even when the laser frequency is on
resonance with the optically allowed electronic transition. Ironically, nuclear spin
diffusion mediated by heavy holes is allowed due to the electron hyperfine interaction
which drives the system to a quasi optical dark state. The longer the system stays in the
dark state the more efficient diffusion becomes. Our results not only provide an
explanation for the experimentally observed low degrees of nuclear spin polarization, but
they also offer an alternative explanation to the results found in
Ref.~\cite{chekhovich2013} since the orientation of nuclear spins cannot be assumed to be
solely defined by the pumping scheme. Finally, we simultaneously propose a simple
experiment aiming at detecting and cancelling the effective heavy hole non-collinear
interaction.

The effective Hamiltonian~\cite{supp},
\begin{equation}
	H = H'_0 + H'_{\mm{L}} + H_{\mm{Z}}^{\mm{nuc}} + H_{\mm{HF},z}^{\ue} +
	H_{\mm{HF},z}^{\uh} + H_{\mm{HF},\mm{nc}}^{\uh},
	\label{eq:Hpump}
\end{equation}
describes the coherent dynamics of the system in the presence of an external magnetic
field oriented along the growth axis of the quantum dot (Faraday geometry) and when the
laser frequency is close to resonance with the transition $\ket{0} \leftrightarrow
\ket{\!\!\downarrow \Uparrow}$ [c.f.  Fig.~\ref{fig:opticalprocesses}]. The Hamiltonian
$H'_0$ describes the evolution of the exciton states,
\begin{equation}
\begin{aligned}
	H'_0 &= \frac{\hbar \Delta}{2}\left(-\ket{0}\bra{0} + \ket{\!\!\downarrow \Uparrow}\bra{\downarrow \Uparrow\!\!}\right)
	+ \left(\frac{\hbar \Delta}{2} + E^{\uparrow \Downarrow}_{\downarrow \Uparrow}\right)
	\ket{\!\!\uparrow \Downarrow}\bra{\uparrow \Downarrow\!\!}\\
	&\phantom{=}
	+ \left(\frac{\hbar \Delta}{2} + E^{\uparrow \Uparrow}_{\downarrow
	\Uparrow}\right) \ket{\!\!\uparrow \Uparrow}\bra{\uparrow \Uparrow \!\!} 
	+ \left(\frac{\hbar \Delta}{2} + E^{\downarrow \Downarrow}_{\downarrow
	\Uparrow}\right)\ket{\!\!\downarrow \Downarrow}\bra{\downarrow \Downarrow \!\!},
\end{aligned}
\label{eq:H0rf}
\end{equation}
where $\Delta$ is the laser detuning and we have 
\begin{equation}
\begin{aligned}
E^{\uparrow \Downarrow}_{\downarrow \Uparrow} &= -\sqrt{\delta_1^2 + g_-^2\mu_{\mm{B}}^2 B^2},\\
E^{\uparrow \Uparrow}_{\downarrow \Uparrow} &= -\delta_0 + \sqrt{\delta_2^2 +
g_+^2\mu_{\mm{B}}^2 B^2} - \sqrt{\delta_1^2 + g_-^2\mu_{\mm{B}}^2 B^2},\\
E^{\downarrow \Downarrow}_{\downarrow \Uparrow} &= -\delta_0 -
\sqrt{\delta_2^2 + g_+^2\mu_{\mm{B}}^2 B^2} - \sqrt{\delta_1^2 + g_-^2\mu_{\mm{B}}^2 B^2}.
\end{aligned}
\label{eq:diffexenerg}
\end{equation}
Here, we have defined $g_+ = g_{\ue} + 3 g_{\uh}$ and $g_- = g_{\ue} - 3 g_{\uh}$ with
$g_{\ue}$ ($g_{\uh}$) the electron (heavy hole) Landé $g$-factor, and $\mu_{\mm{B}}$ is
the Bohr magneton. The coefficients $\delta_0$, $\delta_1$, and $\delta_2$ describe
respectively the fine structure splitting between bright and dark excitons, among bright,
and among dark excitons~\cite{bayer2002}. Since we are considering $\sigma_+$ circularly
polarized light and working in a Faraday geometry, the evolution of $\ket{\!\!\uparrow
\Downarrow}$ is trivial. We can therefore reduce the complexity of the problem by
omitting this state. The laser Hamiltonian reads
\begin{equation}
	H'_{\mm{L}} = \hbar \Omega\left(\ket{0}\bra{\downarrow \Uparrow\!\!} +
	\ket{\!\!\downarrow \Uparrow}\bra{0}\right).
\label{eq:HlaserRWA}
\end{equation}
where $\Omega$ is the Rabi frequency. The nuclear Zeeman Hamiltonian is given by
\begin{equation}
	H_{\mm{n}}^{\mm{Z}} = g_{\mm{n}} \mu_{\mm{n}} B I_z,
	\label{eq:HnZ}
\end{equation}
with $g_{\mm{n}}$ the nuclear Landé $g$-factor and $\mu_{\mm{n}}$ the nuclear Bohr
magneton. The electronic hyperfine Hamiltonian within the homogeneous coupling
approximation is given by,
\begin{equation}
	H_{\mm{HF}}^{\ue} =  H_{\mm{HF},z}^{\ue} + H_{\mm{HF},\perp}^{\ue} = A^{\ue}
	\left(S_z I_z + \frac{1}{2}\left(S_+ I_- + S_- I_+\right)\right).
	\label{eq:hfe}
\end{equation}
Here, $S_z$ ($I_z$) is the electron (nuclear) spin operator in $z$ direction and we have
introduced the ladder operators, $S_\pm = S_x \pm \ui S_y$ and $I_\pm = I_x \pm \ui I_y$.
We denote the average hyperfine coupling constant as $A^{\ue}$, the longitudinal part of
Eq.~\eqref{eq:hfe} by $H_{\mm{HF},z}^{\ue}$, and the transverse one by
$H_{\mm{HF},\perp}^{\ue}$. The effective hyperfine Hamiltonian for heavy holes can be
written as~\cite{supp},
\begin{equation}
\begin{aligned}
	H_{\mm{HF}}^{\uh} &= H_{\mm{HF},z}^{\uh} + H_{\mm{HF},\perp 1}^{\uh} +
	H_{\mm{HF},\perp 2}^{\uh} + H_{\mm{HF},\mm{nc}}^{\uh} \\ 
	& = A^{\uh}_z S^{\uh}_z I_z + A_{\perp 1}^{\uh} S^{\uh}_+ I_- +  A_{\perp 1}^{\uh \ast} S^{\uh}_- I_+ \\
	&\phantom{=} + A_{\perp 2}^{\uh} S^{\uh}_+ I_+ + A_{\perp 2}^{\uh \ast}  S^{\uh}_- 
	I_- + A^{\uh}_{\mm{nc}} S^{\uh}_z I_+ + A^{\uh \ast}_{\mm{nc}} S^{\uh}_z  I_-,
	\label{eq:hfh}
\end{aligned}
\end{equation}
where $S^{\uh}_i$ ($i=z,\pm$) are pseudospin operators for the effective heavy hole
states~\cite{supp}. We use a similar notation to the one introduced in Eq.~\eqref{eq:hfe}
for longitudinal and transverse interactions. We denote the non-collinear term by
$H_{\mm{Hf},\mm{nc}}^{\uh}$.

Here, we follow the procedure of Refs.~\cite{latta2009,hildmann2013} and describe the
contribution of the transverse (flip-flop) terms of Eqs.~\eqref{eq:hfe} and
\eqref{eq:hfh} on the dynamics as a dissipative process. The evolution of the system is
then described by the Lindblad master equation~\cite{lindblad1976}, $\dot{\rho} =
-\frac{\ui}{\hbar} [H,\rho] + \sum_{j=1}^{d^2 -1} ( [L_j \rho, L_j^{\dag}] +
[L_j, \rho L_j^{\dag}])/2$, with $d$ the dimension of the Hilbert space.  Since we only
consider dissipative processes within the electronic subspace, we get by with less
Lindblad operators. We take into account spontaneous emission from the bright exciton
$\ket{\!\!\downarrow \Uparrow}$ and from both (quasi) dark excitons to the ground
state~\cite{chekhovich2013}. These
are respectively described by $L_1 = \sqrt{\Gamma_{\mm{sp}}^{\downarrow \Uparrow}}
\ket{0}\bra{\downarrow \Uparrow\!\!}$, $L_2 =
\sqrt{\Gamma_{\mm{sp}}^{\uparrow \Uparrow}} \ket{0}\bra{\uparrow \Uparrow\!\!}$, and $L_3
= \sqrt{\Gamma_{\mm{sp}}^{\downarrow \Downarrow}} \ket{0}\bra{\downarrow
\Downarrow\!\!}$, where $\Gamma_{\mm{sp}}^j$, $j=\downarrow\Uparrow, \uparrow\Uparrow,
\downarrow\Downarrow$, is the spontaneous decay rate. We describe the nuclear spin
state with its total angular momentum $j$ and its projection along the magnetic field
given by $m$ [c.f. Fig.~\ref{fig:opticalprocesses}]. The Lindblad operators $L_4 =
\sqrt{\Gamma_{\mm{e}}^{\downarrow \Uparrow}} \ket{0,j,m-1}\bra{\downarrow \Uparrow,j,m}$
and $L_5 = \sqrt{\Gamma_{\mm{\uh}}^{\downarrow \Uparrow}} \ket{0,j,m+1}\bra{\downarrow
\Uparrow,j,m}$ respectively describe electron and hole flip-flop processes. The
rates $\Gamma_{\ue}^{\downarrow \Uparrow}$ and $\Gamma_{\uh}^{\downarrow \Uparrow}$ are
calculated with the same method as in Ref.~\cite{hildmann2013}, we find 
\begin{equation}
	\Gamma_{\ue}^{\downarrow \Uparrow} \simeq \frac{\Gamma_{\mm{sp}}^{\uparrow
	\Uparrow}}{4}\abs{\frac{A^{\ue}\sqrt{j(j+1)-m(m-1)}}{E^{\downarrow
	\Uparrow}_{\uparrow \Uparrow}-A^{\ue}(m-\frac{1}{2}) + \frac{3}{2}A_z^{\uh} +
	g_{\mm{n}}\mu_{\mm{n}}B}}^2,
\label{eq:frateelec}
\end{equation}
where we have neglected the contribution coming from $H_{\mm{HF},\perp 2}^{\uh}$ since
$\abs{A_{\perp,2}^{\uh}}/A^{\ue}\ll 1$ and 
\begin{equation}
	\Gamma_{\uh}^{\downarrow \Uparrow} = \frac{\Gamma_{\mm{sp}}^{\downarrow
	\Downarrow}}{4}\abs{\frac{A^{\uh}_{\perp,1}\sqrt{j(j+1)-m(m+1)}}{E^{\downarrow
	\Uparrow}_{\downarrow \Downarrow} + \frac{1}{2}A^{\ue} + 3
	A_z^{\uh}(m+\frac{1}{2}) - g_{\mm{n}}\mu_{\mm{n}}B}}^2.
\label{eq:fratehole}
\end{equation}

\begin{figure*}[t]
\includegraphics[width=2\columnwidth]{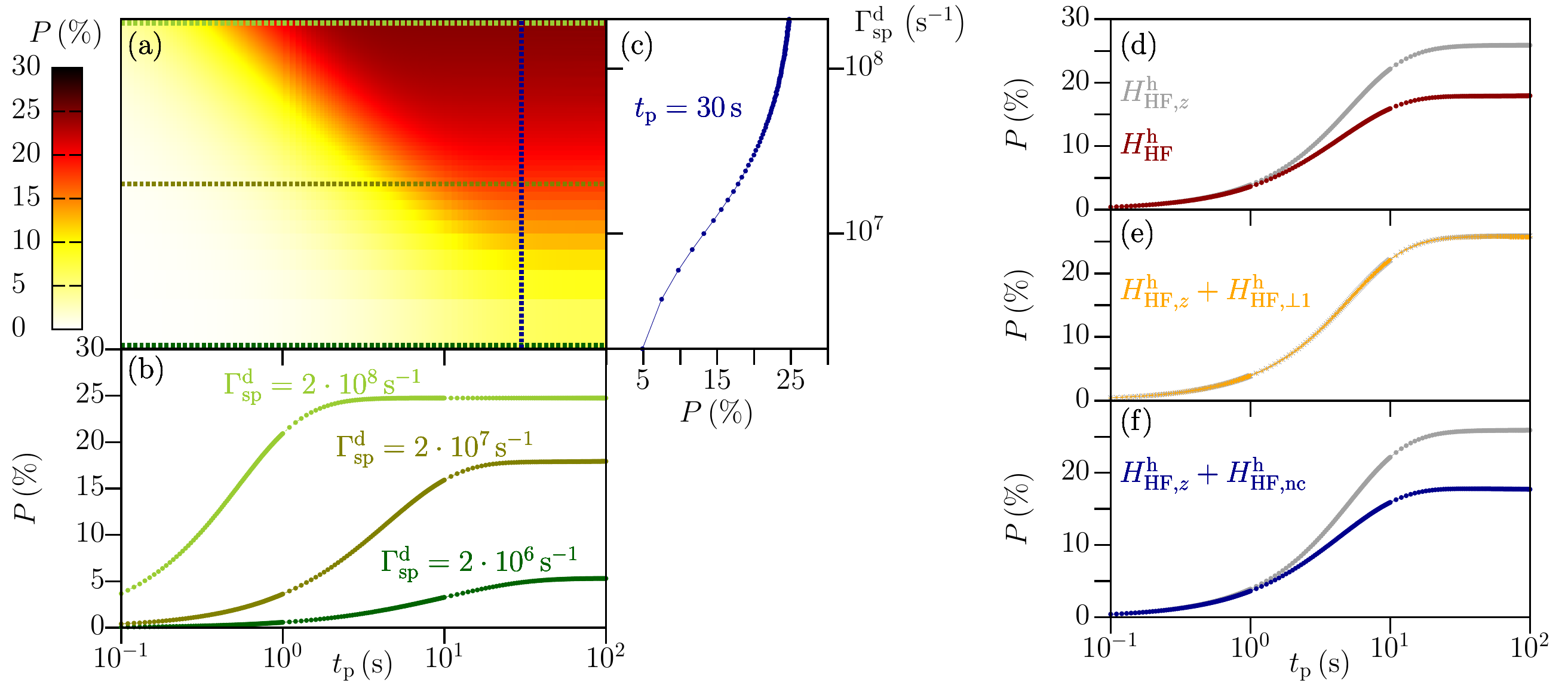}
\caption{(Color online). (a) Nuclear spin polarization $P$ as a function of pumping time
$t_{\mm{p}}$ and spontaneous decay rate $\Gamma_{\mm{sp}}^{\uparrow \Uparrow}\approx
\Gamma_{\mm{sp}}^{\downarrow \Downarrow} = \Gamma_{\mm{sp}}^{\mm{d}}$. Values of other parameters
are given in the main text. (b) Traces taken along $\Gamma_{\mm{sp}}^{\mm{d}} = 2\cdot
10^8,\,2\cdot 10^7,\,\mm{and}\,2\cdot 10^{6}\,\mm{Hz}$. $P$ saturates at
lower values for smaller $\Gamma_{\mm{sp}}^{\mm{d}}$'s. (c) Trace taken along $t_{\mm{p}}
= 30\,\mm{s}$ showing the dependence on $\Gamma_{\mm{sp}}^{\mm{d}}$. (d) Comparison of
$P$ as a function of $t_{\mm{p}}$ between $H_{\mm{HF}}^{\mm{h}}=H_{\mm{HF},z}^{\uh}$
(gray), for which saturation corresponds to formation of a nuclear spin dark state,  and
the full effective Hamiltonian Eq.~\eqref{eq:hfh} (red) for
$\Gamma_{\mm{sp}}^{\mm{d}}=2\cdot 10^7\,\mm{Hz}$.  (e) Same as (d) but compared with
$H_{\mm{HF}}^{\mm{h}}=H_{\mm{HF},z}^{\uh} + H_{\mm{HF},\perp 1}^{\uh}$. The result shows
that heavy hole mediated flip-flop processes are negligible. (f) Same as (d) but compared
with $H_{\mm{HF}}^{\mm{h}}=H_{\mm{HF},z}^{\uh} + H_{\mm{HF},\mm{nc}}^{\uh}$. The heavy
hole hyperfine non-collinear interaction is the origin of the lower values of $P$ since
it leads to an effective nuclear spin diffusion mechanism.}
\label{fig:results_sat}
\end{figure*}
We assume nuclear spins to be initially in a thermal state. This is a reasonable
assumptions even for experiments performed at low temperatures, where the thermal energy
is larger than the nuclear Zeeman, $k_{\mm{B}} T \gg E_{\mm{Z}}^{\mm{nuc}}$, with
$k_{\mm{B}}$ the Boltzmann's constant. Thus, at $t=0$, the nuclear spins are assumed to
be in a fully mixed state. Further assuming spin-1/2 for the nuclei, we
have~\cite{hildmann2013}
\begin{equation}
	\rho_{\mm{nuc}} = \sum_{j,m}\frac{(2j+1)
	N!\left[\Theta(j+m)-\Theta(j-m)\right]}{\left(\frac{N}{2} + j + 1\right)!
	\left(\frac{N}{2} -j\right)! 2^N}\ket{j,m}\bra{j,m},
\label{eq:rhonucinit}
\end{equation}
with $N$ the number of nuclear spins and $\Theta(x)$ is the Heaviside theta function.
The initial electronic state is given by the quantum dot vacuum, i.e. $\rho_{\ue} =
\ket{0}\bra{0}$. Thus, the density matrix describing the whole system at $t=0$ is written
as $\rho = \rho_{\ue} \otimes \rho_{\mm{nuc}}$. 

In Fig.~\ref{fig:results_sat}(a), we present the degree of nuclear spin polarization
$P=\mm{Tr}\left[\rho_{\mm{nuc}}(t_{\mm{p}}) I_z\right]/P_{\mm{max}}$ as a function of
pumping time $t_{\mm{p}}$ and spontaneous decay rate of the dark states
$\Gamma^{\mm{d}}_{\mm{sp}}$, where $P_{\mm{max}}=N/2$. Since the ratio of the dark states
energy is nearly one, $E_{\uparrow \Uparrow}/E_{\downarrow \Downarrow}\approx
1$~\cite{supp}, we have $\Gamma_{\mm{sp}}^{\uparrow \Uparrow} \simeq
\Gamma_{\mm{sp}}^{\downarrow \Downarrow} = \Gamma^{\mm{d}}_{\mm{sp}}$. The calculations
were performed with $\delta_0=5.6213\cdot10^{11}\,\mm{Hz}$, $\delta_1 = \delta_2 =
5.3174\cdot10^{10}\,\mm{Hz}$, $B=8\,\mm{T}$, $g_{\ue}=-0.35$, $g_{\uh}=0.63$, $g_{\mm{n}}
\mu_{\mm{n}} = 3.3\cdot10^{-8}\,\mm{eV/T}$, $\Delta =0\,\mm{Hz}$, $\Omega=2.03\cdot
10^{10}\,\mm{Hz}$, $A^{\ue}=10^8\,\mm{Hz}$, $A_z^{\uh}=-10^7\,\mm{Hz}$, $\abs{A_{\perp
1}^{\uh}}= 3\cdot 10^5\,\mm{Hz}$, $A_{\mm{nc}}^{\uh}= 3\cdot 10^5\,\mm{Hz}$,
$\Gamma_{\mm{sp}}^{\downarrow \Uparrow} = 2\cdot 10^9\,\mm{Hz}$, and $N=30$. We have
cancelled out the imaginary part of $H_{\mm{HF},\mm{nc}}^{\uh}$ by performing a suitable
rotation of angle $\theta$, $U=\exp[\ui \theta I_z]$. We notice that the saturation of
the nuclear polarization depends strongly on the lifetime of the dark states. To
demonstrate clearly this behavior, we present traces taken respectively for different
values of $\Gamma_{\mm{sp}}^{\mm{d}}$ [Fig.~\ref{fig:results_sat}(b)] and for
$t_{\mm{p}}=30\,\mm{s}$ [Fig.~\ref{fig:results_sat}(c)], for which $P$ has reached
saturation. Both of these traces show that $P$ saturates at smaller values for slower
$\Gamma_{\mm{sp}}^{\mm{d}}$. To come to a clear mechanism that explains this result, we
compare $P$ as  function of $t_{\mm{p}}$ for $\Gamma_{\mm{d}}= 10^{7}\,\mm{s^{-1}}$
between an Ising-like ($H_{\mm{HF},z}^\uh$) and other forms of the effective heavy hole
hyperfine Hamiltonian. In Fig.~\ref{fig:results_sat}(d), we compare $P$ obtained with
$H_{\mm{HF},z}^{\uh}$ (gray) to the effective Hamiltonian given by Eq.~\eqref{eq:hfh}
(red). The different values of $P$ at saturation indicate that the relatively small
corrections to the Ising-like hyperfine Hamiltonian influence the dynamics. For
$H_{\mm{HF},z}^{\uh}$, $P$ saturates due to formation of a nuclear spin dark
state~\cite{hildmann2013}. To tell apart the contribution of $H_{\mm{HF},\perp 1}^{\uh}$
and $H_{\mm{HF},\mm{nc}}^{\uh}$, we plot $P$ obtained with $H_{\mm{HF},z}^{\uh}$ (gray)
and $H_{\mm{HF},z}^{\uh} + H_{\mm{HF},\perp 1}^{\uh}$ (orange) in
Fig.~\ref{fig:results_sat}(e), which show that heavy hole flip-flop processes are
irrelevant for the nuclear spin dynamics. In the presence of a large magnetic field and
due to the smallness of $\abs{A_{\perp,1}^{\uh}}$, the heavy hole forbidden relaxation
rate $\Gamma_{\mm{h}}^{\downarrow \Uparrow}$ is too slow compared to electron forbidden
relaxation rate $\Gamma_{\mm{e}}^{\downarrow \Uparrow}$ and to spontaneous emission
$\Gamma_{\mm{sp}}^{\downarrow \Uparrow}$ to have an impact on the dynamics.  Finally, we
verify that the non-collinear interaction is responsible for saturations lower than the
nuclear dark state limit. In Fig.~\ref{fig:results_sat}(f), we show $P$ calculated with
$H_{\mm{HF},z}^{\uh}$ (gray) and $H_{\mm{HF},z}^{\uh} + H_{\mm{HF},\mm{nc}}^{\uh}$
(blue). Our results demonstrate that the heavy hole non-collinear hyperfine interaction
leads to an effective nuclear spin diffusion mechanism that hinders $P$. As it can be
observed from Figs.~\ref{fig:results_sat}(a), (b), and (c), the diffusion becomes more
prominent when the system is hold for a relatively long time in one of the optical dark
states. It has been reported that the oscillator strength for optical dark states is a
hundred to a thousand times smaller than the oscillator strength of bright
states~\cite{chekhovich2013}, which implies $\Gamma_{\mm{sp}}^{\downarrow
\Uparrow}/\Gamma_{\mm{sp}}^{\mm{d}} \approx 100-1000$. Finally, our results indicate that
upon reaching saturation most of the nuclear spin states are still populated and the
system is left in a mixed state. Thus, our findings suggest that there could be an alternative
interpretation of recent experimental results about the sign of the Ising-like
interaction~\cite{chekhovich2013}.  The unexpected shift of the Overhauser field could
simply originate from nuclear spin diffusion, which lowers $P$, when measuring the
spectral position of the optical dark states. 

In the following, we propose a simple experiment to detect and simultaneously
cancel the presence of non-collinear interactions. The idea is to change the orientation
of the external magnetic field to transform the nuclear Zeeman Hamiltonian into,
$H_{\mm{n}}^Z = g_{\mm{n}} \mu_{\mm{n}} B \cos(\varphi) I_z + g_{\mm{n}} \mu_{\mm{n}} B
\sin(\varphi)(I_+ + I_-)/2$, with $\varphi$ the rotation angle. In our coordinate system
the magnetic field has to be rotated around the $y$-axis, i.e. $\varphi$ is the angle
between the $z$-axis and $\Bv$. We solve again a Lindblad master
equation, but with a Hamiltonian that takes into account that $\Bv$ is not necessarily
aligned with the growth axis of the quantum dot. In addition to the trivial change $B \to
B\cos(\varphi) \equiv B_z$ in Eqs.~\eqref{eq:Hpump}, \eqref{eq:frateelec}, and
\eqref{eq:fratehole} as well as the discussed modification of the nuclear Zeeman
Hamiltonian, we also need to take into account that misalignment of $\Bv$ leads to mixing
of bright and dark excitons via $H_{\mm{bd}} = g_{\ue} \mu_{\mm{B}} B \sin(\varphi) (S_+
+ S_-)/4 + g_{\uh}^{xx} \mu_{\mm{B}} B \sin(\varphi)(S_+^{\uh} + S_-^{\uh})/4$, with
$g_{\uh}^{xx} \simeq g_{\uh}/10$~\cite{bayer2000,godden2012} the heavy hole Landé
$g$-factor along the $x$-axis. We also add to the dissipative part of the Lindblad
equation spontaneous relaxation from $\ket{\!\!\uparrow \Downarrow}$ to the ground state
with rate $\Gamma_{\mm{sp}}^{\uparrow \Downarrow}$ and two non-conserving nuclear spin
relaxation mechanisms. These are described by $L_6 = \sqrt{\Gamma_{\mm{sp}}^{\uparrow
\Downarrow}} \ket{0}\bra{\uparrow \Downarrow\!\!}$, $L_7 = \sqrt{\Gamma_{\ue}^{\uparrow
\Downarrow}}\ket{0, j, m+1}\bra{\uparrow \Downarrow, j, m}$, and $L_8 =
\sqrt{\Gamma_{\uh}^{\uparrow \Downarrow}}\ket{0, j, m-1}\bra{\uparrow \Downarrow, j, m}$
with
\begin{equation}
	\Gamma_{\ue}^{\uparrow \Downarrow} \simeq \frac{\Gamma_{\mm{sp}}^{\downarrow
	\Downarrow}}{4}\abs{\frac{A^{\ue}\sqrt{j(j+1)-m(m+1)}}{E^{\uparrow
	\Downarrow}_{\downarrow \Downarrow}+A^{\ue}(m+\frac{1}{2}) +
	\frac{3}{2}A_z^{\uh} - g_{\mm{n}}\mu_{\mm{n}}B_z}}^2,
	\label{eq:frateelec2}
\end{equation}
and 
\begin{equation}
	\Gamma_{\uh}^{\uparrow \Downarrow} = \frac{\Gamma_{\mm{sp}}^{\uparrow
	\Uparrow}}{4}\abs{\frac{A^{\uh}_{\perp 1}\sqrt{j(j+1)-m(m-1)}}{E^{\uparrow
	\Downarrow}_{\uparrow \Uparrow} + \frac{1}{2}A^{\ue} +
	\frac{3}{2}A_z^{\uh}(m-\frac{1}{2}) + g_{\mm{n}}\mu_{\mm{n}}B_z}}^2.
	\label{eq:fratehole2}
\end{equation}

\begin{figure}
\includegraphics[width=\columnwidth]{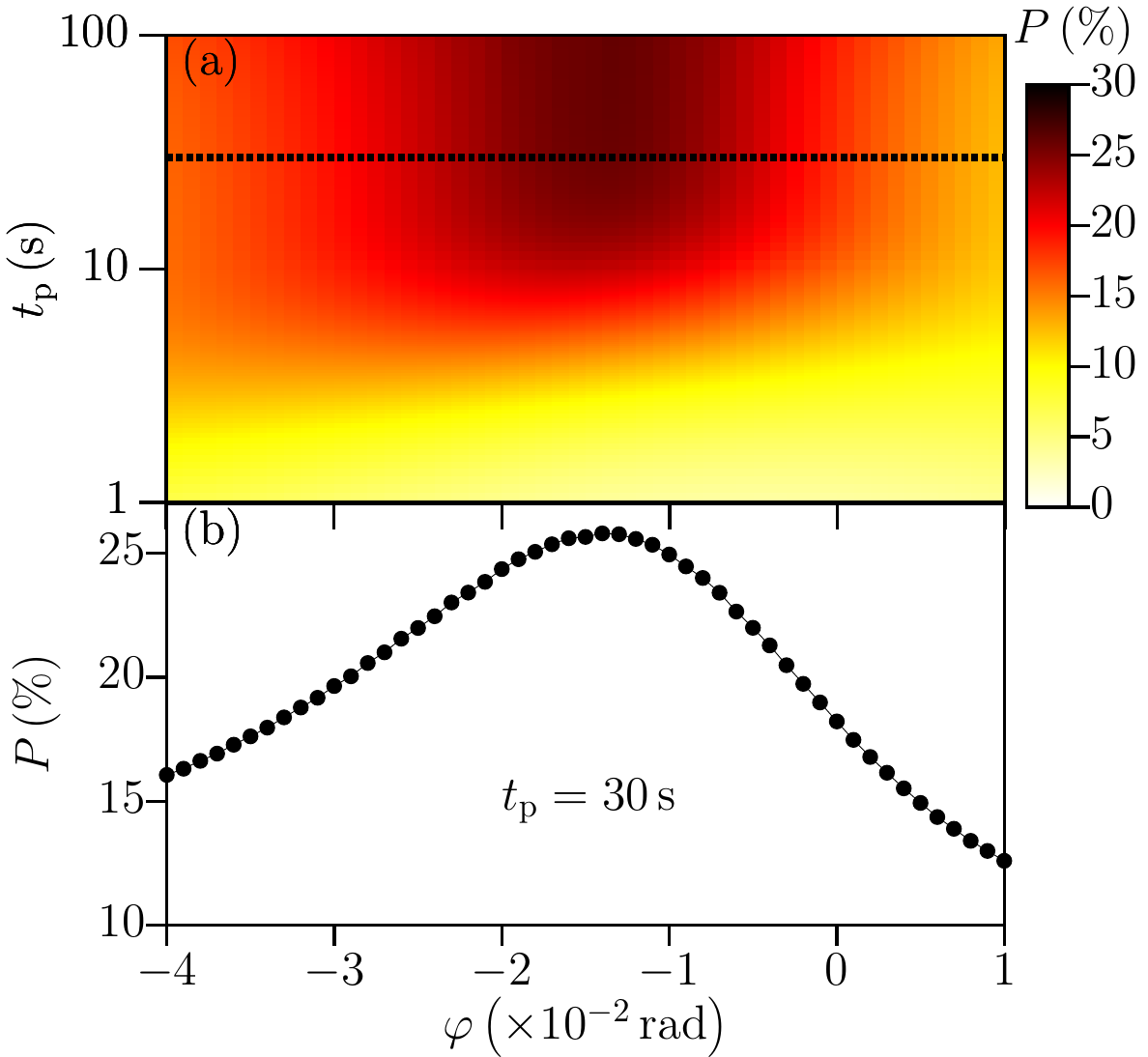}
\caption{(Color online). (a) Nuclear spin polarization $P$ as a function of $t_{\mm{p}}$
and $\varphi$ (angle between the external magnetic field $\Bv$ and the $z$-axis) for
$\Gamma_{\mm{sp}}^{\mm{d}}= 2\cdot 10^7\,\mm{s^{-1}}$. (b) Trace taken along
$t_{\mm{p}}=30\,\mm{s}$. The effect of the non-collinear interaction can be cancelled by
orienting the magnetic field opposite to the effective Overhauser field defined by the
hyperfine non-collinear Hamiltonian.}
\label{fig:results_phi}
\end{figure}

In Fig.~\ref{fig:results_phi}(a), we plot the nuclear spin polarization $P$ as a function of
$t_{\mm{p}}$ and $\varphi$. We use the same set of parameters as before and
$\Gamma_{\mm{sp}}^{\mm{d}} = 10^7\,\mm{s^{-1}}$. As for the optical dark state, we have
$E_{\downarrow \Uparrow}/E_{\uparrow \Downarrow} \approx 1$ which allows us to write
$\Gamma_{\mm{sp}}^{\uparrow \Downarrow} \simeq \Gamma_{\mm{sp}}^{\downarrow \Uparrow} =
\Gamma_{\mm{sp}}^{\mm{b}}$. The results show that the non-collinear heavy hole hyperfine
interaction is fully cancelled at $\varphi \simeq -0.014\,\mm{rad}$, for which we
retrieve the saturation limit set by the nuclear spin dark state. In
Fig.~\ref{fig:results_phi}(b), we show a trace taken for $t_{\mm{p}} = 30\,\mm{s}$.

In conclusion, we have shown that the effective heavy hole hyperfine interaction via
non-collinear terms influences nuclear spin dynamics. In particular, we have shown how to
experimentally detect and cancel the effects of such interaction. We expect the described
effects to be stronger when considering an inhomogeneous hyperfine Hamiltonian since the
statistical weight of the states contributing the most to $P$ are not
suppressed~\cite{hildmann2013}. Moreover, when trying to cancel the heavy hole
non-collinear interaction, a series of maximums should be observed as a function of the
rotation angles. Each maximum corresponds to a different nuclear species. This also
implies that none of the maximums correspond to the limit set by the formation of a
nuclear dark state.

We are thankful to J. Hildmann and R. J. Warburton for fruitful discussions. We
acknowledge funding from the Swiss NF, NCCR QSIT, and $\mm{S}^3$ NANO.

\end{document}


\title{Supplementary Material:~Nuclear Spin Diffusion Mediated by Heavy Hole Hyperfine
Non-Collinear Interactions}

\author{Hugo Ribeiro}
\affiliation{Department of Physics, University of Basel, Klingelbergstrasse 82, CH-4056
Basel, Switzerland}

\author{Franziska Maier}
\affiliation{Department of Physics, University of Basel, Klingelbergstrasse 82, CH-4056
Basel, Switzerland}

\author{Daniel Loss}
\affiliation{Department of Physics, University of Basel, Klingelbergstrasse 82, CH-4056
Basel, Switzerland}

\maketitle

\section{Effective heavy hole states in a self-assembled quantum dots}
\label{sec:effHHstates}
In this section, we introduce the $4\times4$ $\bm{k}\cdot\bm{p}$ Hamiltonians describing
valence band states in zincblende semiconductors confined to strained quantum dots
with applied external magnetic field.  Furthermore, we calculate the hybridized
lowest-energy eigenstates of the heavy-hole subsystem which are subject to
light-hole mixing~\cite{fischer2008,fischer2010,maier2012}.

The Hamiltonians~\cite{winkler2003} are written in terms of the spin-$3/2$ matrices $J_i$,
$i=x,y,z$, which are given in a basis of angular momentum eigenstates $\ket{J, M}$. Here,
the heavy hole band corresponds to
$M=\pm3/2$ and the light hole band to $M=\pm1/2$. We choose the basis of the bulk
Hamiltonians to be $\{\ket{u_{3/2}},\ket{u_{1/2}},\ket{u_{-1/2}},\ket{u_{-3/2}}\}$, where
the $\ket{u_{M}}$ are  products of $P$-symmetric orbital angular momentum
eigenstates ($\ket{P_{\pm}}, \ket{P_{z}}$) and spin states
($\ket{\uparrow},\ket{\downarrow}$)~\cite{winkler2003}:  $\ket{u_{\pm3/2}} = \mp
1/\sqrt{2}\ket{P_{\pm}}\ket{\uparrow,\downarrow}$ and $\ket{u_{\pm1/2}} =
1/\sqrt{6}(\ket{2 P_{z},P_{-}}\ket{\uparrow}\mp\ket{P_{+}, 2 P_{z}}\ket{\downarrow})$.

In the bulk semiconductor, the heavy and light hole states are given by the Luttinger-Kohn Hamiltonian
\begin{equation}
\begin{aligned}
H_k &= -\frac{\hbar^2}{2 m_0}\left(\gamma_1 k^2 -2 \gamma_2 \left((J_x^2-1/3
J^2)k_x^2+\mathrm{c.p.}\right)\right) + \frac{\hbar^2}{2 m_0}4\gamma_3\left(\{J_x, J_y\}\{k_x,
k_y\}+\mathrm{c.p.}\right)\\
&\phantom{=} +  \frac{2}{\sqrt{3}}C_k \left(\{J_x, J_y^2-J_z^2\}k_x+\mathrm{c.p.}\right),
\end{aligned}
\label{eq:LKham}
\end{equation}
where we have defined $\{A,B\} = (AB+BA)/2$, $\mm{c.p.}$ denotes cyclic permutation,
$\hbar k_i = -i \hbar \partial_i$, $i=x,y,z$, is the momentum operator, $k^2 =
k_x^2+k_y^2+k_z^2$ and $J^2 = J_x^2+J_y^2+J_z^2$.  The Luttinger parameters are given by
$\gamma_l$, $l=1,2,3$ and $C_k$ is a consequence of the spin-orbit interaction with
higher bands. We denote the diagonal part of $H_k$ by $H_{k,0}$.

We include strain by taking into account 
\begin{equation}
H_{{\varepsilon}} = D_d \,\mathrm{Tr}\varepsilon+2/3 D_u \left((J_x^2-1/3
J^2)\varepsilon_{xx}+\mathrm{c.p.}\right) \left(C_4 (\varepsilon_{yy}-\varepsilon_{zz})J_x k_x
+\mathrm{c.p.}\right),
\label{eq:Hstrain}
\end{equation}
where we have only considered diagonal elements of the strain
tensor $\varepsilon$, $\varepsilon_{ii}$, $i=x,y,z$. $D_d$ and $D_u$ denote vector potentials and the constant $C_4$ is
defined in Ref.~\cite{trebin1979}.  We refer to the diagonal,
$kv$-independent part of $H_{{\varepsilon}}$ as $H_{{\varepsilon},0}$.

A magnetic field, $\Bv = \nabla \times \Av = (0,0,B_z)$, pointing along the growth
direction of the quantum dot, is included by adding two more terms to the
Hamiltonian\cite{ivchenko1998,kiselev1998}. The first term is found by replacing
$\kv \rightarrow \kv+ e \Av$ in $H_k+H_{\varepsilon}$ in a semi-classical manner.
This yields the implicit magnetic field dependence given by the vector potential
$\Av$. We keep only terms linear in $\Av$ and define
\begin{equation}
H_{mc} = e \Av \cdot \vv,
\end{equation}
where $\vv = \partial (H_k+H_{\varepsilon})/\partial \kv$ is the velocity operator.
We note that proper operator ordering is still enforced.  The second term is the magnetic
interaction term
\begin{equation}
H_{B} = -2 \mu_B \left(\kappa J_z B_z + q J_z^3 B_z\right),
\end{equation}
where $\kappa$ is the isotropic and $q$ the anisotropic part of the hole $g$ factor.

We model a flat, cylindric quantum dot by choosing a three-dimensional harmonic confinement potential, 
\begin{equation}
V_{c} = \left(\begin{array}{cccc}
  V_{c,\text{HH}} &0&0&0\\
  0& V_{c,\text{LH}} &0&0\\
  0&0& V_{c,\text{LH}}&0\\
  0&0&0& V_{c,\text{HH}}
          \end{array}
\right), 
\end{equation}
with
\begin{equation}
V_{c,b}(\mathbf{r}) = -\frac{1}{2}m_{b,\perp} \omega_{b,\perp}^2 z^2 - \frac{1}{2} m_{b,\|} \omega_{b,\|}^2(x^2+y^2),
\end{equation}
where $b = \mm{HH}, \mm{LH}$ detones heavy and light hole. We assume the origin of the coordinate system
to be loaced at the center of the quantum dot. The confinement energies $\omega_{b,
\|}=\hbar/(m_{b, \|}L_{b}^2)$ and $\omega_{b, \perp}=\hbar/(m_{b, \perp}a_{b}^2)$ are
defined by the confinement lengths $L_{b}$ and $a_{b}$ ($L_{b}\gg a_{b}$).  The
corresponding effective masses in the single bands are given by $m_{\mm{HH/LH}, \perp}=
m_0/(\gamma_1 \mp 2 \gamma_2)$ and $m_{\mm{HH/LH}, \|}= m_0/(\gamma_1 \pm \gamma_2)$.

The quantum dot states are then described by
\begin{equation}
	H_{\mm{qd}} = H_k+H_{\varepsilon} +H_{mc}+H_B + V_{c},
\end{equation}
and the used parameter values can be found in Tab.~\ref{tab:parameters}.  A realistic
strain configuration for cylindric InAs quantum dots can be found in
Ref.~\cite{tadic2002}. We divide $H_{\mm{qd}}$ into a leading order term $H_{\mm{qd},0}
= H_{k,0}+H_{{\varepsilon},0}+ V_{c}$ and a perturbation $H_{\mm{qd},1}$, where $H_{k,0}$
and $H_{{\varepsilon},0}$ denote the diagonal terms of $H_{k}$ and $H_{{\varepsilon}}$.
We directly map the Hamiltonian $H_{k,0}+ V_{c}$ onto a three dimensional anisotropic
harmonic oscillator whose eigenenergies $E_{b}$ in band $b$ are given by
\begin{eqnarray}
E_{b} &=& -\hbar \omega_{b,\perp} (n_z+\frac{1}{2})-\hbar \omega_{b,\|} (n_x+n_y+1).
\end{eqnarray}
The associated eigenfunctions are the usual three dimensional harmonic oscillator
eigenfunctions [see e.g.~\cite{sakurai1994}]
$\phi_{b,\nvs}(\rv)$, where $\nv = (n_x,n_y,n_z)$ is a vector of the associated
quantum numbers. We choose the basis states of $H_{\mm{qd},0}$ to be products of type 
\begin{equation}
\ket{\Psi_{m_j}^{\nv}} = \phi_{b}^{\nvs}(\rv)\ket{u_{m_j}}.
\label{eq:basisstate}
\end{equation}
We are interested in the two lowest-energy states in the heavy hole band,
$\ket{\Psi_{3/2}^{\nvs_{0}}}$ and $\ket{\Psi_{-3/2}^{\nvs_{0}}}$ ($\nv_{0}=(0,0,0)$). We
decouple these states from the higher energy states of $H_{\mm{qd}}$ by a
Schrieffer-Wolff transformation~\cite{schrieffer1966}, $\tilde{H}_{\mm{qd}} =
\ue^{-S}H_{\mm{qd}}\ue^{S}$. We perform the transformation up to second order in
$(H_{\mm{qd},1})_{ij}/\Delta E \ll 1$, where $(H_{\mm{qd},1})_{ij}$ denotes the matrix
elements coupling $\ket{\Psi_{3/2}^{\nvs_{0}}}$ and $\ket{\Psi_{-3/2}^{\nvs_{0}}}$ to the
higher energy states and $\Delta E$ is the energy splitting between the associated
subspaces. The approximate lowest energy eigenstates are given by
\begin{equation}
\ket{\tilde{\Psi}_{\pm3/2}^{\nvs_{0}}} =\mathcal{N}_{\pm3/2,\nvs_{0}} \sum_{M,\nvs}
c_{M,\nvs} \ket{\Psi_{M}^{\nvs}},
\end{equation}
where $\mathcal{N}_{\pm3/2,\nvs_{0}}$ denotes the normalization. The
coefficients $c_{M,\nvs}$ depend on quantum dot parameters, such as confinement,
strain, dot material, and on the external magnetic field. 
\begin{table}[h]
\begin{tabular}{lcddclldd}
\cline{1-4}\cline{6-9}\\[-0.42cm]\cline{1-4}\cline{6-9}
		&& \text{GaAs\hspace{-0.3cm}}	& \text{InAs\hspace{-0.3cm}}	&\mbox{\hspace{0.3cm}}& 	&& \text{GaAs\hspace{-0.3cm}} & \text{InAs\hspace{-0.3cm}}\\	\cline{1-4}\cline{6-9}
$\gamma_1$	&& 6.85	& 20.40							&& $C_k$ &[eV\AA]	& -0.0034  & -0.0112 \\
$\gamma_2$	&& 2.10	& 8.30							&& $D_d$ &[eV]	& -1.16\cite{vurgaftman2001}  & -1.0\cite{vurgaftman2001} \\
$\gamma_3$	&& 2.90	& 9.10							&& $D_u$ &[eV]	& 3.0 & 2.7 \\
$\kappa$	&& 1.1\cite{traynor1995}	& 7.68\cite{traynor1995}	&& $C_4$ &[eV\AA]	& 6.8\cite{silver1992} & 7.0\cite{silver1992} \\
$q$		&& 0.01\cite{mayer1991} 	& 0.04\cite{lawaetz1971}	&& $a$ & [\AA]	& 5.65 & 6.05 \\\cline{1-4}\cline{6-9}\\[-0.42cm]\cline{1-4}\cline{6-9}
\end{tabular}
\caption{List of the used material parameters. All parameters were taken from
	Ref.~\cite{winkler2003} if not stated otherwise. \label{tab:parameters}}
\end{table}

\section{Effective hyperfine interactions of heavy holes}

In this section, we calculate the coupling between the effective lowest-energy states
$\ket{\tilde{\Psi}_{3/2}^{\nvs_{0}}}$ and $\ket{\tilde{\Psi}_{-3/2}^{\nvs_{0}}}$ via
the hyperfine interaction in InAs (GaAs) quantum dots. We follow the procedure outlined
in Refs.~\cite{fischer2008,fischer2010}.

The hyperfine interaction Hamiltonians for an electron located at $\rv$ in the field
of a nucleus at $\Rv_k$ consists of three terms, the Fermi contact hyperfine
interaction ($H_{\mm{HF}}^{\mm{c}}$), the anysotropic hyperfine interaction
($H_{\mm{HF}}^{\mm{a}}$) and the coupling
of the orbital angular momentum to the nuclear spin ($H_{\mm{HF}}^{\mm{L}}$). The three
terms read ($\hbar=1$)~\cite{stoneham1972}
\begin{align}
	H_{\mm{HF}}^{\mm{c}} &= \frac{\mu_0}{4 \pi} \frac{8 \pi}{3} \gamma_S \gamma_{j_k}
	\Sv \cdot \Iv_k \delta(\rv_{k}),\\
	H_{\mm{HF}}^{\mm{a}} &= \frac{\mu_0}{4 \pi} \gamma_S \gamma_{j_k} \frac{3 (\nv_k \cdot 
  	\Sv) (\nv_k \cdot \Iv_k) - \Sv \cdot \Iv_k}
  	{r_{k}^3 (1+d/r_{k})},\\
   	H_{\mm{HF}}^{\mm{L}} &= \frac{\mu_0}{4 \pi} \gamma_S \gamma_{j_k} \frac{\Lv_k \cdot 
 	 \Iv_k} {r_{k}^3 (1+d/r_{k})}.
\end{align}
Here $\gamma_S=2\mu_B$ and $\gamma_{j_k}=g_{j_k} \mu_N$ with the Bohr (nuclear) magneton
$\mu_B$ ($\mu_N$) and $g_{j_k}$ is the nuclear $g$ factor of isotopic species $j_k$. We
have defined the relative coordinate of the electron with respect to the $k$th nucleus by
$\rv_{k} = \rv - \Rv_k$, its relative direction by $\nv_k = \rv_{k}/ r_{k}$, and $d
\simeq Z \times 1.5 \times 10^{-15} \, \mathrm{m}$, and $Z$ is the charge of the nucleus.
The spin and orbital angular-momentum operators are denoted by $\Sv$ and $\Lv_k = \rv_k
\times \pv$.

The Bloch states are written in terms of orbital angular momentum and spin states as introduced in Sec.\
\ref{sec:effHHstates}.  We approximate the orbital angular momentum eigenstates as linear combinations of
atomic eigenfunctions~\cite{gueron1964} $u_j(\rv) = \alpha \psi_{In,
Ga}^{nlm}(\rv + \dv/2) + \sqrt{1-\alpha^2}\psi_{As}^{4lm}(\rv-\dv/2)$.  Here,
$\alpha$ is the electron distribution between the two atoms and $\psi^{nlm}(\rv) =
R_{nl}(r)Y_l^m(\vartheta, \varphi)$ are hydrogenic eigenfunctions with quantum numbers
$n$, $l$, and $m$.  The radial part of the wavefunction depends on the effective central
charge $Z_{\text{eff}}$ of the nuclei where we use values for free atoms
\cite{clementi1963, clementi1967}. The position of the hole with respect to a nuclei
located in the Wigner-Seitz cell at $\pm\dv/2$ is denoted $\rv\pm\dv/2$, where $\dv=
a_{\mm{In(Ga)As}}(1,1,1)/4$ is the In(Ga)-As bonding vector
defined by the respective lattice constant $a_{\mm{In(Ga)As}}$. The bonding  character
of the valence band is expressed by the $+$ sign, and $\int_{\mm{WS}} \ud^3 r
|u_{m_j}(\rv)|^2=2$ enforces normalization~\cite{coish2009}. Here, the subscript
WS indicates that the integral is evaluated over the Wigner-Seitz cell. For the numerical
evaluation of the integral we consider a Wigner-Seitz cell as a sphere of radius one half of the
As-As atom distance, centered in the middle of the bond connecting In(Ga) with As.

To calculate the hyperfine interaction of different basis states with all nuclear spins,
we have to evaluate $\sum_k \bra{\Psi_{M}^{\nvs}(\rv)} H_{\mm{HF}}^{\mm{c}} +
H_{\mm{HF}}^{\mm{a}} + H_{\mm{HF}}^{\mm{L}}
\ket{\Psi_{M'}^{\nvs'}(\rv)}$.  We note that the envelopes
$\phi_{b}^{\nvs}(\rv)$ vary slowly across a single Wigner-Seitz cell, whereas the dominant
part of the hyperfine interaction is given within a single Wigner-Seitz cell and the long-ranged
part of the hyperfine interactions spreading across the boundaries of the Wigner-Seitz cell leads
only to minor corrections\cite{fischer2008}. Hence we can write the hyperfine
interaction as $\sum_k (\phi_{b}^{\nvs}(\rv_k))^{*} \phi_{b'}^{\nvs'}(\rv_k)
\bra{u_{M}^k} H_{\mm{HF}}^{\mm{c}} + H_{\mm{HF}}^{\mm{a}} + H_{\mm{HF}}^{\mm{L}}
\ket{u_{m_j'}^k}_{\mm{WS}}$. We numerically evaluate the
matrix elements $\bra{u_{M}^k} H_{\mm{HF}}^{\mm{c}} +
H_{\mm{HF}}^{\mm{a}} + H_{\mm{HF}}^{\mm{L}} \ket{u_{M'}^k}_{\mm{WS}}$ and write them
in matrix form using the basis
$\{\ket{u_{3/2}^k},\ket{u_{1/2}^k},\ket{u_{-1/2}^k},\ket{u_{-3/2}^k}\}$. By averaging our
results over the nuclear abundance $\nu_i$ of each atomic species $i$, we find
\begin{equation}
H_{\mm{HF}} \approx A_{\text{hf}}^h \left(
\begin{array}{cccc}
 - I_z & -0.52 I_{-} & 0 & 0 \\
 -0.52 I_{+} & -0.46 I_z & -0.65 I_{-} & 0 \\
 0 & -0.65 I_{+} & 0.46 I_z & -0.52 I_{-} \\
 0 & 0 & -0.52 I_{+} &  I_z \\
\end{array}
\right)\end{equation}
where $A_{\text{hf}}^h\sim 1.58 \mbox{ $\mu$eV}  (1.21 \mbox{ $\mu$eV})$ is the InAs
(GaAs) hyperfine coupling constant.  
Using $H_{\mm{HF}}$ we can derive an effecive hyperfine Hamiltonian
for the lowest-energy states $\ket{\tilde{\Psi}_{\pm3/2}^{\nvs_0}}$,
We find for an InAs quantum dot subject to
$B_z = 5 \mbox{T}$ with $\varepsilon_{xx} = \varepsilon_{yy} = -\varepsilon_{zz} =
-0.06$, $L_{\text{LH}} = 1.5 L_{\text{HH}}$, and $a_{\text{LH}} = 1.5 a_{\text{HH}}$
\begin{equation}
\begin{aligned}
	H_{\mm{HF}}^{\uh} &= A^{\uh}_z S^{\uh}_z I_z + A_{\perp,1}^{\uh} S^{\uh}_+
	I_- + A_{\perp,1}^{\uh \ast}  S^{\uh}_- I_+
	+ A_{\perp,2}^{\uh} S^{\uh}_+ I_+ + A_{\perp,2}^{\uh \ast}  S^{\uh}_-
	I_- + A^{\uh}_{\mm{nc}} S^{\uh}_z I_+ + A^{\uh \ast}_{\mm{nc}} S^{\uh}_z  I_-\\
	&\phantom{=} + A^{\uh}_{\mm{ncs}} S_+ I_z +  A^{\uh \ast}_{\mm{ncs}} S_- I_z,
\end{aligned}
\label{eq:hfhf}
\end{equation}
with $A^{\uh}_z\simeq -7.89\cdot~10^{-7}\,\mm{eV}$, $A_{\perp,1}^{\uh} \simeq -
4.13\cdot~10^{-10} \ui$, $A_{\perp,2}^{\uh} \simeq -
6.38\cdot~10^{-10}$, $A_{\mm{ncs}}^{\uh} \simeq - 5.03\cdot~10^{-10} (1+\ui)$, and
$A_{\mm{nc}}^{\uh} \simeq - 1.58\cdot~10^{-9} (1-\ui)$.

\section{Effective Hamiltonian for optical nuclear spin pumping via bright excitons}

The Hamiltonian describing the dynamics of the system is given by
\begin{equation}
	H(t)=H_0 + H_{\mm{L}}(t) + H_{\mm{Z}}^{\mm{nuc}} + H_{\mm{HF}}^{\ue} + H_{\mm{HF}}^{\uh}.
\label{eq:Htot}
\end{equation}
The Hamiltonian $H_0$ describes the free evolution of the exciton states, $H_{\mm{L}}(t)$
is the time-dependent laser Hamiltonian, $H_{\mm{Z}}^{\mm{nuc}}$ describes the nuclear
Zeeman interaction, and $H_{\mm{HF}}^{\ue}$ and $H_{\mm{HF}}^{\uh}$ are
the hyperfine Hamiltonians for electrons and holes. We have 
\begin{equation}
	H_0 = E_{\downarrow \Uparrow} \ket{\!\!\downarrow \Uparrow}\bra{\downarrow \Uparrow\!\!} 
	+ E_{\uparrow \Downarrow} \ket{\!\!\uparrow \Downarrow}\bra{\uparrow \Downarrow\!\!} 
	+ E_{\uparrow \Uparrow} \ket{\!\!\uparrow \Uparrow}\bra{\uparrow \Uparrow \!\!} 
	+ E_{\downarrow \Downarrow} \ket{\!\!\downarrow \Downarrow}\bra{\downarrow
	\Downarrow \!\!},
	\label{eq:HO}
\end{equation}
where 
\begin{equation}
\begin{aligned}
	E_{\downarrow \Uparrow} &= E_{\mm{X}} + \gamma_2 B^2 + \frac{\delta_0}{2} +
	\frac{1}{2}\sqrt{\delta_1^2 + (g_{\ue} - 3 g_{\uh})^2\mu_{\mm{B}}^2 B^2}\\
	E_{\uparrow \Downarrow} &= E_{\mm{X}} + \gamma_2 B^2 + \frac{\delta_0}{2} -
	\frac{1}{2}\sqrt{\delta_1^2 + (g_{\ue} - 3 g_{\uh})^2\mu_{\mm{B}}^2 B^2}\\
	E_{\uparrow \Uparrow} &= E_{\mm{X}} + \gamma_2 B^2 -\frac{\delta_0}{2} 
	+ \frac{1}{2}\sqrt{\delta_2^2 + (g_{\ue} + 3 g_{\uh})^2\mu_{\mm{B}}^2 B^2}\\
	E_{\downarrow \Downarrow} &= E_{\mm{X}} + \gamma_2 B^2 -\frac{\delta_0}{2} 
	- \frac{1}{2}\sqrt{\delta_2^2 + (g_{\ue} + 3 g_{\uh})^2\mu_{\mm{B}}^2 B^2}.
\end{aligned}
\label{eq:energy_excitons}
\end{equation}
Here, $E_{\mm{X}}$ is the energy required to excite an exciton (band-gap energy),
$\gamma_2 B_z^2$ is the diamagnetic shift~\cite{bayer1998}, with $\gamma_2$ the
diamagnetic constant and $B$ the $z$-component of the external magnetic field,
$\mu_{\mm{B}}$ is the Bohr magneton, and $g_{\ue}$ ($g_{\uh}$) is the electron (hole) Landé
$g$-factor. The fine structure of excitons~\cite{bayer2002} results in an energy
splitting $\delta_0$ between bright and dark exciton subspaces, $\delta_1$ between bright
excitons, and $\delta_2$ between dark excitons. 

The nuclear Zeeman Hamiltonian is given by
\begin{equation}
	H_{\mm{n}}^{\mm{Z}} = g_{\mm{n}} \mu_{\mm{n}} B I_z,
	\label{eq:HnZ}
\end{equation}
with $g_{\mm{n}}$ the nuclear Landé $g$-factor and $\mu_{\mm{n}}$ the nuclear Bohr
magneton. The time-dependent laser Hamiltonian can be written as
\begin{equation}
	H_{\mm{L}}(t) = \hbar\Omega\,\left(\ue^{\ui\omega_{\mm{L}} t}\ket{0}\bra{\downarrow \Uparrow\!\!}
	+ \ue^{-\ui \omega_{\mm{L}} t}\ket{\!\!\downarrow \Uparrow}\bra{0}\right),
\label{eq:Hlaser}
\end{equation}
where $\Omega$ and $\omega_{\mm{L}}$ are respectively the Rabi and laser frequencies. The
electronic hyperfine Hamiltonian within the homogeneous coupling approximation is given by,
\begin{equation}
	H_{\mm{HF}}^{\ue} = A^{\ue} \left(S_z I_z + \frac{1}{2}\left(S_+ I_- + S_-
	I_+\right)\right).
\label{eq:hfe}
\end{equation}
The average hyperfine coupling strength is denoted by $A^{\ue}$. $S_z$ and $I_z$ describe
respectively the $z$-component of the electron spin and total nuclear spin. We have also
introduced electronic and nuclear spin ladder operators, $S_\pm = S_x \pm \ui S_y$ and
$I_\pm = I_x \pm \ui I_y$. The effective hyperfine Hamiltonian for heavy holes can be
written,
\begin{equation}
	H_{\mm{HF}}^{\uh} = A^{\uh}_z S^{\uh}_z I_z + A_{\perp,1}^{\uh} S^{\uh}_+
	I_- + A_{\perp,1}^{\uh \ast}  S^{\uh}_- I_+
	+ A_{\perp,2}^{\uh} S^{\uh}_+ I_+ + A_{\perp,2}^{\uh \ast}  S^{\uh}_-
	I_- + A^{\uh}_{\mm{nc}} S^{\uh}_z I_+ + A^{\uh \ast}_{\mm{nc}} S^{\uh}_z  I_-,
	\label{eq:hfh}
\end{equation}
where $S^{\uh}_i$ ($i=z,\pm$) are pseudospin operators for the heavy hole states. 

To get rid of the explicit time-dependence in $H_{\mm{L}}$, we perform a similarity
transformation $H\to \ue^{\ui \xi t}(H-\xi)\ue^{-\ui \xi t}$, with $\xi = E_{\downarrow
\Uparrow} + \hbar \Delta$ and $\Delta$ is the laser detuning. This transformation leaves
both hyperfine and nuclear Zeeman Hamiltonians unchanged, whereas $H_0$ becomes
\begin{equation}	
\begin{aligned}
	H_0 \to H_0' &= \frac{\hbar \Delta}{2}\left(-\ket{0}\bra{0} +
	\ket{\!\!\downarrow \Uparrow}\bra{\downarrow \Uparrow\!\!}\right) +
	\left(\frac{\hbar \Delta}{2} + E^{\uparrow \Downarrow}_{\downarrow
	\Uparrow}\right) \ket{\!\!\uparrow \Downarrow}\bra{\uparrow \Downarrow\!\!}\\
	&\phantom{=} + \left(\frac{\hbar \Delta}{2} + E^{\uparrow \Uparrow}_{\downarrow
	\Uparrow}\right) \ket{\!\!\uparrow \Uparrow}\bra{\uparrow \Uparrow \!\!} +
	\left(\frac{\hbar \Delta}{2} + E^{\downarrow \Downarrow}_{\downarrow
	\Uparrow}\right)\ket{\!\!\downarrow \Downarrow}\bra{\downarrow \Downarrow \!\!},
\end{aligned}
\label{eq:H0rf}
\end{equation}
with $E^i_j = E_i - E_j$. For the laser Hamiltonian, after performing the
similarity transformation, we also make the rotating wave approximation and neglect fast
counter-oscillating terms, we find ($H_{\mm{L}}\to H'_{\mm{L}}$),
\begin{equation}
	H_{\mm{L}}(t) \to H'_{\mm{L}} = \hbar \Omega\left(\ket{0}\bra{\downarrow \Uparrow\!\!} +
	\ket{\!\!\downarrow \Uparrow}\bra{0}\right).
	\label{eq:HlaserRWA}
\end{equation}

The total Hamiltonian in the rotating frame and within the rotating wave approximation is
thus given by
\begin{equation}
	H' = H'_0 + H'_{\mm{L}} + H_{\mm{n}}^{\mm{Z}} + H_{\mm{HF}}^{\ue} +
	H_{\mm{HF}}^{\uh}.	
	\label{eq:HtotRWA}
\end{equation}

We further eliminate the flip-flop terms of Hamiltonians Eqs.~\eqref{eq:hfe}
and \eqref{eq:hfh} by applying a Schrieffer-Wolf transformation~\cite{schrieffer1966,
bravyi2011} to $H'$,
\begin{equation}
	H' \to \tilde{H} = \ue^S H' \ue^{-S}= \sum_{j=0}^{\infty} \frac{\left[S,H'\right]^{(j)}}{j!},
\label{eq:SW}
\end{equation}
where we have defined the recursive relation 
\begin{equation} 
\begin{aligned} 
	\left[S,H'\right]^{(0)} &= H' , \\
	\left[S,H'\right]^{(1)} &= \left[S,H'\right] , \\ 
	\left[S,H'\right]^{(j)}
	&= \left[S, \left[S,H'\right]^{(j-1)}\right].  
\end{aligned}
\label{eq:defad} 
\end{equation}

We perform the transformation with 
\begin{equation}
\begin{aligned}
	S &= \frac{1}{2} \left(\frac{A^{\ue}}{E^{\uparrow \Uparrow}_{\downarrow
	\Uparrow}} I_- \ket{\!\!\uparrow \Uparrow}\bra{\downarrow \Uparrow} +
	\frac{A^{\ue}}{E^{\uparrow \Downarrow}_{\downarrow \Downarrow}} I_- 
	\ket{\!\!\uparrow \Downarrow}\bra{\downarrow \Downarrow\!\!} +
	\frac{A^{\ue}}{E^{\downarrow \Downarrow}_{\uparrow \Downarrow}} I_+
	\ket{\!\!\downarrow \Downarrow}\bra{\uparrow \Downarrow\!\!} +
	\frac{A^{\ue}}{E^{\downarrow \Uparrow}_{\uparrow \Uparrow}} I_+
	\ket{\!\!\downarrow \Uparrow}\bra{\uparrow \Uparrow\!\!} \right.\\
	&\phantom{= \frac{1}{2} (} +
	\frac{A^{\uh}_{\perp,1}}{E^{\downarrow \Uparrow}_{\downarrow
	\Downarrow}}I_- \ket{\!\!\downarrow \Uparrow}\bra{\downarrow \Downarrow\!\!} +
	\frac{A_{\perp,1}^{\uh}}{E^{\uparrow \Uparrow}_{\uparrow \Downarrow}} I_-
	\ket{\!\!\uparrow \Uparrow}\bra{\uparrow \Downarrow\!\!} +
	\frac{A_{\perp,1}^{\uh \ast}}{E^{\uparrow \Downarrow}_{\uparrow \Uparrow}}I_+
	\ket{\!\!\uparrow \Downarrow}\bra{\uparrow \Uparrow\!\!} +
	\frac{A_{\perp,1}^{\uh \ast}}{E^{\downarrow \Downarrow}_{\downarrow \Uparrow}}
	I_+ \ket{\!\!\downarrow \Downarrow}\bra{\downarrow \Uparrow\!\!}\\
	&\phantom{= \frac{1}{2} (} \left. +
	\frac{A^{\uh}_{\perp,2}}{E^{\downarrow \Uparrow}_{\downarrow
	\Downarrow}}I_+ \ket{\!\!\downarrow \Uparrow}\bra{\downarrow \Downarrow\!\!} +
	\frac{A_{\perp,2}^{\uh}}{E^{\uparrow \Uparrow}_{\uparrow \Downarrow}} I_+
	\ket{\!\!\uparrow \Uparrow}\bra{\uparrow \Downarrow\!\!} +
	\frac{A_{\perp,2}^{\uh \ast}}{E^{\uparrow \Downarrow}_{\uparrow \Uparrow}}I_-
	\ket{\!\!\uparrow \Downarrow}\bra{\uparrow \Uparrow\!\!} +
	\frac{A_{\perp,2}^{\uh \ast}}{E^{\downarrow \Downarrow}_{\downarrow \Uparrow}}
	I_- \ket{\!\!\downarrow \Downarrow}\bra{\downarrow \Uparrow\!\!}\right),
\end{aligned}
\label{eq:swHpump}
\end{equation}
and obtain by keeping only terms that are first order in $A^{\ue}$ or $A^{\uh}_i$ the
following Hamiltonian,
\begin{equation}
\begin{aligned}
	\tilde{H} &= H_0' + H'_{\mm{L}} + H_{\mm{Z}}^{\mm{nuc}} + A^{\ue} S_z I_z + A^{\uh}_z
	S_z^{\uh} I_z + A^{\uh}_{\mm{nc}} S_z^{\uh}\left(I_+ + I_-\right)\\ 
	&\phantom{=} + 
	\frac{\hbar \Omega}{2} \frac{A^{\ue}}{E^{\uparrow \Uparrow}_{\downarrow \Uparrow}} \left(
	I_+ \ket{0}\bra{\uparrow \Uparrow \!\!} + I_- \ket{\!\!\uparrow \Uparrow}\bra{0}\right) +
	\frac{\hbar \Omega}{2 E^{\downarrow \Downarrow}_{\downarrow \Uparrow}} 
	\left(A^{\uh}_{\perp,1} I_- \ket{0}\bra{\downarrow \Downarrow \!\!} + A^{\uh \ast}_{\perp,1}I_+ \ket{\!\!\downarrow
	\Downarrow}\bra{0}\right)\\ 
	&\phantom{=}+ 
	\frac{\hbar \Omega}{2 E^{\downarrow
	\Downarrow}_{\downarrow \Uparrow}} \left(A^{\uh}_{\perp,2}  I_+
	\ket{0}\bra{\downarrow \Downarrow \!\!} + A^{\uh \ast}_{\perp,2}  I_- \ket{\!\!\downarrow
	\Downarrow}\bra{0}\right).
\end{aligned}
\label{eq:Hfinal}
\end{equation}
The three last terms in Eq.~\eqref{eq:Hfinal} describe hyperfine assisted spin-forbidden
optical transitions. These terms are only relevant when the laser detuning is close to
resonance either with $\ket{\!\!\uparrow \Uparrow}$ or $\ket{\!\!\downarrow
\Downarrow}$~\cite{hildmann2013}. Thus, for $\Delta$ close to zero the coherent dynamics of excitons
and nuclear spins can be described by 
\begin{equation}
	H = H_0' + H'_{\mm{L}} + H_{\mm{Z}}^{\mm{nuc}} + A^{\ue} S_z
	I_z + A^{\uh}_z S_z^{\uh} I_z + A^{\uh}_{\mm{nc}} S_z^{\uh}\left(I_+ + I_-\right).
	\label{eq:Hdet0}
\end{equation}